\documentclass[review]{elsarticle}

\usepackage{upgreek}

\usepackage[hyphens]{url}

\usepackage{amsmath}

\usepackage[T1]{fontenc}

\usepackage{array}
\usepackage{multicol}

\usepackage{graphicx}
\graphicspath{{./pdf/}{./jpeg/}{./images/}}
   \DeclareGraphicsExtensions{.pdf,.jpeg,.jpg,.png,.PNG}

\usepackage{hyperref}

\journal{Computers and Security}

\begin{document}

\begin{frontmatter}

\title{Network Investigation Methodology for BitTorrent Sync: A Peer-to-Peer Based File Synchronisation Service %\tnoteref{mytitlenote}
}

%\tnotetext[mytitlenote]{This is an extended publication from a paper entitled ``BitTorrent Sync: Network Investigation Methodology'', published as part of the proceedings of the 9th International Conference on Availability, Reliability and Security (ARES 2014). All packet analysis has been updated to reflect the changes in the latest production (1.4) and alpha (2.0) releases.}

\author{Mark Scanlon, Jason Farina, M-Tahar Kechadi}
\address{School of Computer Science and Informatics,\\
University College Dublin,
Belfield, Dublin 4, Ireland\\ Email: mark.scanlon@ucd.ie, jason.farina@ucdconnect.ie, tahar.kechadi@ucd.ie \newline \newline Final accepted version available: \url{http://dx.doi.org/10.1016/j.cose.2015.05.003}}



\begin{abstract}
%The volume of personal information and data most Internet users find themselves amassing is ever increasing and the fast pace of the modern world results in most requiring instant access to their files. Millions of these users turn to cloud based file synchronisation services, such as Dropbox, Microsoft OneDrive, Apple iCloud and Google Drive, to enable ``always-on'' access to their most up-to-date data from any computer or mobile device with an Internet connection. The prevalence of recent articles covering various invasion of privacy issues and data protection breaches in the media has caused many to review their online security practices with their personal information. To provide an alternative to cloud based file backup and synchronisation, BitTorrent Inc. released a cloudless file backup and synchronisation service, named BitTorrent Sync. Its popularity rose dramatically since release, reaching over ten million active users by August 2014. This paper outlines a number of scenarios where the network investigation of the service may prove invaluable as part of a digital forensic investigation. An investigation methodology is proposed outlining the required steps involved in retrieving digital evidence from the network and the results from a proof of concept investigation are presented.
High availability is no longer just a business continuity concern. Users are increasingly dependant on devices that consume and produce data in ever increasing volumes. A popular solution is to have a central repository which each device accesses after centrally managed authentication. This model of use is facilitated by cloud based file synchronisation services such as Dropbox, OneDrive, Google Drive and Apple iCloud. Cloud architecture allows the provisioning of storage space with ``always-on'' access. Recent concerns over unauthorised access to third party systems and large scale exposure of private data have made an alternative solution desirable. These events have caused users to assess their own security practices and the level of trust placed in third party storage services. One option is BitTorrent Sync, a cloudless synchronisation utility provides data availability and redundancy. This utility replicates files stored in shares to remote peers with access controlled by keys and permissions. While lacking the economies brought about by scale, complete control over data access has made this a popular solution. The ability to replicate data without oversight introduces risk of abuse by users as well as difficulties for forensic investigators. This paper suggests a methodology for investigation and analysis of the protocol to assist in the control of data flow across security perimeters. 
\end{abstract}


\end{frontmatter}

%\linenumbers

\section{Introduction}

Applications such as Evernote and Dropbox leverage the decreasing cost of hard disk storage seen in Infrastructure as a Service providers, e.g., Amazon S3, to provide data storage on the cloud to home users and businesses alike. The main advantage of services such as Dropbox, Google Drive, Microsoft OneDrive (formally SkyDrive) and Apple iCloud to the end user is that their data is stored in a virtual extension of their local machine with no direct user interaction required after installation. It is also backed up by a fully distributed data-centre architecture that would be completely outside the financial reach of the average consumer. Their data is available anywhere with Internet access and is usually machine agnostic so the same data can be accessed on multiple devices without any need to re-format partitions or wasting space by creating multiple copies of the same file for each device. Some services such as Dropbox, also have offline client applications that allow for synchronisation of data to a local folder for offline access.

As Internet accessibility continues to become more commonplace and allows for increasingly faster access, it is not unexpected that many utilities that are intended for general use will aid in the perpetration of some variety of cybercrime. One attribute that is highly desirable by those contemplating illegal activities is the notion of anonymity and data security -- especially the ability to keep data secure transfer secure from inspection while in transit. BitTorrent Sync (also referred to as BTSync, BitSync and bsync) is a file replication utility that would seem to serve exactly this function for the user. Designed to be server agnostic, the protocol is built on already popular and widespread technologies that would not seem out of place in any network activity log. 

Each of the aforementioned consumer focused services can be categorised as cloud synchronisation services. This means that while the data is synchronised between user machines, a copy of the data is also stored remotely in the cloud. In recent headline news, much of this data is easily available to governmental agencies without the need of a warrant or just cause. BTSync provides the same synchronisation functionality (without the cloud storage aspect) and provides a similar level of data availability. The service has numerous desirable attributes for any Internet user \cite{bitsync}:

\begin{itemize}
\item{Compatibility and Availability} -- Clients are built for most common desktop and mobile operating systems, e.g., Windows, Mac OS, Linux, BSD, Android and iOS.
\item{Synchronisation Options} -- Users can choose whether to sync their content over a local network or over the Internet to remote machines with no requirement for scripting or schedule management making this an accessible technology compared to existing options such as RSYNC.
\item{No Limitations or Cost} -- Most cloud synchronisation services provide a free tier offering a small amount of storage and subsequently charge when the user outgrows the available space. BTSync eliminates these limitations and costs. The only limitation to the volume of storage and speed of the service is down to the limitations of the synchronised users machines.
\item{Automated Backup} -- Like most competing products, once the initial install and configuration is complete, the data contained within specified folders is automatically synchronised between machines.
\item{Decentralised Technology} -- All data transmission and synchronisation takes place solely in a Peer-to-Peer (P2P) fashion, based on the BitTorrent file sharing protocol.
\item{Encrypted Data Transmission} -- While synchronising data between computers, the data is encrypted using RSA encryption. Under the BTSync API, developers can also enable remote file storage encryption \cite{bitsyncapi}. -- This could result in users storing their data on untrusted remote locations for the purposes of data redundancy and secure off site backup.
\item{Proprietary Technology} The precise protocol and operation of the technology is not documented by the developer. There is debate over whether security through obscurity or peer code evaluation, i.e., open source, is better. Some enterprise security policies prohibit the use of open source applications as a result of the source code being open to inspection by those looking for flaws in the implementation. From the point of view of the consumer, BitTorrent Inc. have stated that they will not give access to traffic to any LEA without due process and the bespoke protocol makes casual eavesdropping or crawling less likely.
\end{itemize}

As a result of these attributes, BTSync has grown to become a popular alternative to cloud based synchronisation services. Less than a year after its release, the active user base had grown to over one million by November 2013, doubling to two million by December 2013 \cite{bitsyncstats}, and to over ten million users by August 2014 \cite{bitsync14stats}. Due to this rapid growth and popularity the service will undoubtedly be of interest to both law enforcement officers and digital forensics investigators in future investigations. Like many other file distribution technologies, this interest may be centred around recovery of the data itself, proof of the modification of data or evidence of data distribution and enumeration of the recipients.

While BTSync is based on the same technology as BitTorrent for the transfer of files, the intention of the application is quite different. This results in a change of users' behaviours, as well as a necessary change in the assumptions an investigator should make. BitTorrent is designed to be a one-to-many data dissemination utility. The uploader usually does not care about the identity of the downloader and a single seeder can deliver data to a large number of unique peers over the life of the torrent file. Data integrity and transfer speed take precedence over privacy of data in transit. 
%In this regard BitTorrent can be seen as a stack of pamphlets left on a table for anyone to take as they pass by.
BTSync on the other hand, is designed to be a secure data replication protocol for making a faithful replica of a data set on a remote machine. Data integrity is still highly prised but data privacy is now the top priority and speed-through-dispersion is sacrificed as a result. The files can only be read by users specifically given access to the repository. The advertisement of data availability is completely scalable by the owner with options ranging from restricting access to known IP addresses through to registration with a centralised tracker. Given the nature of the application, users are much more likely to know the operator of the remote site (this does not apply to secrets advertised online though that could be a point of commonality that would not necessarily have existed for pure BitTorrent clients).

%\cite{zhu2013let}, \cite{stolen2013using}  and \cite{lareida2013box2box}. In 2012, \citet{klumpp2012file} outlined blah blah blah.

\subsection{Aim and Contribution of this Work}

The aim of this work is to provide a reference for digital investigators discovering the use of BitTorrent Sync in an active investigation. However, it is hoped that the analysis presented may be of use to security personnel looking to detect and control the use of this protocol within their perimeter. 

To accommodate these goals this work presents an analysis of the protocol and its network interaction. Activities undertaken to perform a synchronisation are presented and described at the packet level in order to facilitate both post mortem traffic analysis and to enable the development of feature based detection rules and deep packet inspection for Network Intrusion Detection Systems (NIDS) or firewall appliances.

The contribution of this work presents a suggested a network investigation methodology for BitTorrent Sync, outlined in Section~\ref{methodology}. This methodology includes recommendations for the investigation of a number of hypothetical scenarios where BTSync could be used to aid in criminal or illicit activities. Legitimate usage of the system, e.g., backup and synchronisation, group modification, data transfer between systems, etc., may itself be of interest to an investigation. However, the technology may also be suitable in the aid of a number of potential scenarios of interest such as industrial espionage, copyright infringement, sharing of illicit images of children, etc., outlined in greater detail in Section~\ref{btsync:usecase}. This work also documents each of the observed packets sent and received during regular operation of BTSync. Finally, the results from two digital forensic investigations of the service are outlined in Section~\ref{proofofconcept} and Section~\ref{investigation} respectively.

%%%%%%%%%%%%%%%%%%%%%%%%%%%%%%%%%%%%%%%%%%%%%%%%%%%%%%%%%%%%%%%%%%%%%%

\section{Background}
\label{background}

%%%%%%%%%%%%%%%%%%%%%%%%%%%%%%%%%%%%%%%%%%%%%%%%%%%%%%%%%%%%%%%%%%%%%%

In order to gain an understanding of how BTSync functions, one must first understand the technologies upon which it is built. The application is a product built by BitTorrent Inc. (the creators and maintainers of the eponymous file-sharing protocol). As a result, the technologies used by the regular BitTorrent protocol and BTSync are developed using a similar premise. This section provides a brief overview of the required background information and outlined the key differences between the two applications.

\subsection{BitTorrent File Sharing Protocol}
\label{bittorrent}

The BitTorrent protocol is designed to easily facilitate the distribution of files to a large number of downloaders with minimal load on the original file source \cite{cohen2008bittorrent}. This is achieved through the downloaders uploading their completed parts of the entire file to other downloaders. A BitTorrent swarm is made up of both seeders (peers with complete copies of the content shared in the swarm), and leechers (peers who are downloading the content and may have none or some of the content). Due to BitTorrent's ease of use and minimal bandwidth requirements, it lends itself as an ideal platform for the unauthorised distribution of copyrighted material. The unauthorised distribution of copyrighted material typically commences with a single original source sharing large sized files to many downloaders.

\subsubsection{Bencoding}
\label{bencoding}
Bencoding is a method of notation for storing data in an array list. The main advantage of bencoding is that it avoids the pitfalls of system-byte order requirements (such as big-endian or little-endian), which can cause issues for cross platform communication between applications. The datagram packet can easily be converted to a human readable UTF-8 encoded sequence of \texttt{key:value} pairs. Indicative \texttt{key:value} pairings are presented in Table \ref{tab:bencoding}.

\begin{table}
\caption{BTSync Packet Bencoding Fields}

\makebox[\linewidth]{
    \begin{tabular}{|l|p{2.9in}|}
\hline
\textbf{Key} & \textbf{Explanation}  \\ \hline
        d:      & Marks the start of a dictionary     \\ \hline
        l:      & List start, the start of a list of field:value pairs in an array. Lists are terminated with an ``e'' \\ \hline
        la:      & local Address IP:Port in Network-Byte Order      \\ \hline
        ea:     & External Address IP:Port in Network-Byte Order\\ \hline
        m:      & Message Type Header, e.g., ping      \\ \hline
        peer:      & [Peer ID]      \\ \hline
        share:      & [Share ID]      \\ \hline
        nonce:    &  16-byte nonce for key exchange between peers negotiating data exchange \\ \hline
        e:      & Marks the end of a dictionary or list      \\ \hline
    \end{tabular}
}
\label{tab:bencoding}
\end{table}

The value for any pair is stored as a sequence of-bytes with the exception of integer values. Associated with the integer indicating keys, bencoding uses the lowercase ``i'' to indicate the start of an integer value, which is also terminated with a lowercase ``e''.

\subsubsection{Active Peer Discovery}
\label{discovery}

Each BitTorrent client must be able to identify a list of active peers in the same swarm who have at least one piece of the content and is willing to share it, i.e., identify a peer that has an available open connection and has the bandwidth available to upload. By the nature of the implementation of the protocol, any peer that wishes to partake in a swarm must be able to communicate and share files with other active peers. BitTorrent provides a number of methods available for peer discovery. There are a number of methods that a BitTorrent client can use in an attempt to discover new peers who are in the swarm outlined below

\begin{enumerate}
\item Tracker Communication -- BitTorrent trackers maintain a list of seeders and leechers for each BitTorrent swarm they are currently supporting \cite{cohen2003incentives}. Each BitTorrent client will contact the tracker intermittently throughout the download of a particular piece of content to report that they are still alive on the network and to download a short list of new peers on the network.
\item Peer Exchange (PEX) -- As set out in the standard BitTorrent specification, there is no intercommunication between peers of different BitTorrent swarms besides data transmission. Peer Exchange is a BitTorrent Enhancement Proposal (BEP) whereby when two peers are communicating (sharing the data referenced by a torrent file), a subset of their respective peer lists are shared during the communication.
\item Distributed Hash Tables (DHT) -- Many BitTorrent clients, such as Vuze and $\upmu$Torrent contain implementations of a common distributed hash table as part of the standard client features. The common DHT maintains a list of each active peer using the corresponding clients and enables cross-swarm communication between peers. Each known peer active in swarms with DHT contributors is added to the DHT. The mainline BitTorrent DHT protocol (also used by BTSync), is based on the Kademlia protocol. Regular BitTorrent file-sharing users and BTSync users contribute to the update and maintenance of the DHT. 
The DHT provides an entirely decentralised approach aiding in the discovery of new peers sharing particular pieces of content. The Kademlia DHT structures its ID space as a tree \cite{li2005comparing}. The distance between two keys in the ID space is their ``exclusive or'' (\texttt{xor}). Each user in the DHT generates a unique key that is used for identification when connecting to the DHT. The piece of the DHT that each peer stores is related to this \texttt{xor} calculation. Those peer IDs that are closest to the key, e.g., a torrent's \texttt{info\_hash}, are responsible for facilitating lookups for those keys. The same DHT responsible for regular BitTorrent file-sharing is also responsible for maintaining a lookup for BTSync shared content. In this scenario, the key used is based on the public read-only key generated for each shared folder in BTSync.

While a DHTs decentralised nature results in a much more resilient service compared to server based tracker, it also results in it be vulnerable to certain attacks, as outlined in greater details in Sit et. al's 2002 paper \cite{sit2002security}.

\item Local Peer Discovery (LPD) -- This is enabled by checking the ``Search LAN'' option in most BitTorrent client's application preferences. When enabled the application will announce its availability to potential local peers using multicast packets. Once a client on the network receives a multicast packet, that client will check its current list of shares to see if a match is found. Is a match it found, that peer will respond to the origin of the request offering to synchronise the content.
\end{enumerate}

%\subsection{File Sync/Cloud Storage/SaaS}
%\label{filesync}
%
%Dropbox metadata sync \cite{Zhang:2013:BTR:2534861.2534863}

\subsubsection{Downloading of Content through BitTorrent}
\label{downloading}

To commence the download of the content in a particular BitTorrent swarm, a metadata \texttt{.torrent} file or a corresponding magnet universal resource identifier (URI) must be acquired from a BitTorrent indexing website. This file/URI is then opened using a BitTorrent client, which proceeds to identify other active peers sharing the specific content required. The client application then attempts to connect to several active members and downloads the content piece by piece. Each BitTorrent swarm is built around a single piece of content which is determined through a unique identifier based on a SHA-1 hash of the file information contained in this UTF-8 encoded metadata file/URI, e.g., name, piece length, piece hash values, length and path.

\subsection{BitTorrent Sync}
\label{bysync}

%\begin{figure}[!t]
%\centering
%\includegraphics[width=0.5\textwidth]{BTStore}
%\caption{BitTorrent Sync}
%\label{btstore}
%\end{figure}

%It is here that the differences and similarities with BTSync can be seen most clearly........

BTSync is a file replication utility created by BitTorrent Inc. and released as a private alpha in April 2013 \cite{bitsync}. It is not a cloud backup solution, nor necessarily intended as any form of off-site storage. Any data transferred using BTSync resides in whole files on at least one of the synchronised devices. This makes the detection of data much simpler for digital forensic purposes as there is no distributed file system, redundant data block algorithms or need to contact a cloud storage provider to get a list of all traffic to or from a container using discovered credentials. The investigation remains an examination of the local suspect machine. However, because BTSync uses DHT to transfer data there is also no central authority to manage authentication or log data access attempts. A suspect file found on a system may have been downloaded from one or many sources and may have been uploaded to one or more recipients. Additionally while the paid services offer up to 1TB of storage (Amazon S3 paid storage plan), the free versions which are much more popular with home users cap at approximately 10GB. BTSync is limited only by the size of the folder being set as a share. Another concern for any investigation into BTSync folders is that unless the system being examined is the owner/originator of the folder being shared, it is quite possible that any files present were downloaded without prior knowledge of their content or nature. Before v2.0, BTSync had no built in content preview facility in its protocol, it merely blindly synchronises from host to target without any selection process available to the user. In v2.0, an option was added to the preferences for each folder that allows the user to only synchronise file titles as a zero byte place holder file. If the file is selected the content of the file is downloaded. An update to the link descriptor in v1.4 allows users to get an approximation of the share size at the time of joining.

\subsubsection{Keys}
\label{keys}

\begin{figure}
\centering
\includegraphics[width=0.8\textwidth]{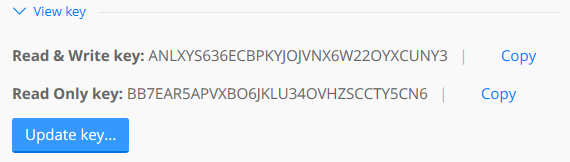}
\caption{Keys (formerly secrets) are generated at share provision. The ability to view the keys is not available in v2.0}
\label{fig:14keys}
\end{figure}

The ``secrets'' used as part of the original release of BTSync were renamed as ``keys'' in v1.4. The structure has not been changed however and still consists of a 33 character human readable string consisting of a Base32 encoded string generated when the folder was first provisioned. This Base32 pattern is then prepended with a single letter indicating its nature. Keys are the unique identifiers used by the BTSync service to differentiate between shared folders. In order for the 20-byte keys to be human readable, they are displayed using Base32 encoding \cite{bitsync}. BTSync facilitates the generation of three categories of secrets for the sharing of data contained within specific folders, as can be seen in Figure~\ref{fig:14keys}.

The initial Read \& Write (RW) key is still generated using \texttt{CryptoApi} on Windows based systems (this is downloaded as part of the installation process if it is not installed already). This RW key is the equivalent of the original ``master secret'' in that, if it is shared then the receiving party has an equal level of access to the share as the original owner including the ability to delete content and add new content that will be replicated to any synchronising peer whether downstream or of equal rank. 

From this initial RW key, a Read Only Key (RO) is generated automatically for sharing. As can be seen in Figure~\ref{fig:14keys}, these are the only two keys readily available to the user. However, these are not the only keys available for use. BTSync defines six standard keys of which three can be generated using the default installation of the desktop client. These keys are identified by their prepended letter as follows:

\begin{itemize}
\item {}[A] This is the RW key generated at the time the share is provisioned. This key gives the user full control over the share contents.
\item {}[B] This identifies the Read Only key and can be used to create a child, or downstream, peer that can only replicate share contents from another peer. Any changes made to share contents, including deletion, will invalidate the file changed and prevent any further replication actions for that particular file in the future, or until the share is re-provisioned on that client (or the share's \texttt{*.db} file is altered but this may cause the entire share to be deemed invalid).
\item {} [C] The C type key is a read only one-use key that is discarded after its first use. This key can be generated from either type A or type B keys and is used primarily in the distribution of other keys.
\item {}[D] Generated through the use of the Sync API, this type of key allows read \& write access to encrypted shares.
\item {}[E] A read only key capable of replicating data from type D encrypted shares and decrypting the contents. This key is calculated form the type D key and so is not possible using that standard BTSync v1.4 or v2.0 installation. 
\item {} [F] Encrypted Read Only key capable of replicating data from an encrypted share but unable to decrypt the share contents. This type of key can be used to store data in an encrypted state on a remote, untrusted, system and still provide authenticity and availability.
\end{itemize}

Older versions of these, such as the `R' prepended read only key of v0.x are still usable but are no longer generated by the application. As with the earlier BTSync versions, a user may also generate his or her own key that has been Base64 encoded. As a result, these default prepended identification letters cannot always be taken as an definite indicator of the access level granted by a key before it has been applied.

The Keys outlined above need never necessarily be shared publicly, i.e., any user can create a number of keys solely for his personal use across his different machines. Depending on the level of access the user wishes to give to a third-party, he can give the corresponding key to any other user through regular one-to-one communication methods (e-mail, instant messaging, social networking, SMS, etc.). If public distribution is desirable, there are a number of public online avenues for BTSync users to share secrets with each other (e.g. \texttt{www.btsynckeys.com}, \texttt{http://www.reddit.com/r/btsecrets}, among others). Version 1.4 presents a change to the method of sharing a link with a peer that has been modified further in v2.0. In v1.4, a user can still view the RW and RO key of a share and can copy this key and send it via any medium to the remote device. Using this method, the remote device user adds a new share and inputs the key causing the share to automatically query a tracker (if this option is left enabled) for the location of remote peers hosting a share matching the applied key. An alternative to this method was added to the client and works as follows:

\begin{figure}
\centering
\includegraphics[width=0.8\textwidth]{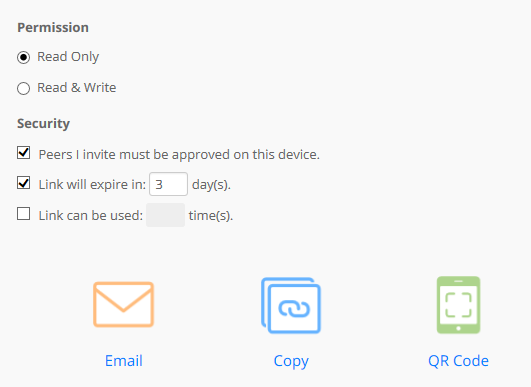}
\caption{Key Sharing is Now Managed from within the Application with Optional Restrictions}
\label{fig:usershare14}
\end{figure}

In the application the user that currently has access to the share (the owner) can select the option to provision the share to another user (a peer which can be a different person or a remote system under the control of the owner), as depicted in Figure~\ref{fig:usershare14}, and is presented with a choice of restrictions and methods presented as options.

\noindent \textbf{Permissions}
\begin{itemize}
\item{Read Only (default)}
\item{ Read \& Write}
\end{itemize}
\textbf{Security Options}
\begin{itemize}
\item{Invited participants must be approved -- the owner will receive notification in the application that a peer wishes to share the resource. The Device ID of the remote peer will be presented and the owner can accept or reject their membership. This option is enabled by default.}
\item{Expiration date -- the link to the share will only remain active for a set number of days from the time it is generated. This option is enabled by default and the time limit is set to three days, but can be changed to any number of days the owner inputs.}
\item{Number of uses -- this option allows the owner to limit the number of times a link can be used to join a share. This is set to off by default.}
\end{itemize}

The link generated by this process is presented as \texttt{https://link.getsync.com/[URLoptions]}, where the URL options are each separated by an ampersand. For example a link shared from v1.4 for a folder called \texttt{winhex} with no expiry or usage limitation would present as \url{https://link.getsync.com/#f=winhex&sz=35E5&s=XIQSFD2MCDPS2QKITWKJROJ2VUSV2YNA&i=CKKR3V2BBM7MXIOTPU3XWK55JBUFWG3EY&p=CALSNMDGCZZAUQXBXEIR6Q57UMTVOSFI&e=1431277452} where:
\begin{itemize}
\item{\#f=(folder name of the share in plain text)}
\item{sz=(approximate size of the share contents)}
\item{s=(the shareID of the folder encoded in Base32)}
\item{i=(a one time key used to provide access to the real key, this changes every time the link for the folder is generated)}
\item{p=(PeerID of the peer performing the server role in the upcoming key exchange)}
\item{e=(the expiry timestamp of the link if it is set, if it is not set this item will not be present in the link)}
\item{v=(the version of the client. This is only present in the v2.0 client and is not optional)}
\end{itemize}

This URL can be copied to the system clipboard, sent via email (the email option will open the default mail application on the system) or converted to a QR code for scanning by a mobile device.

\begin{figure}
\centering
\includegraphics[width=0.8\textwidth]{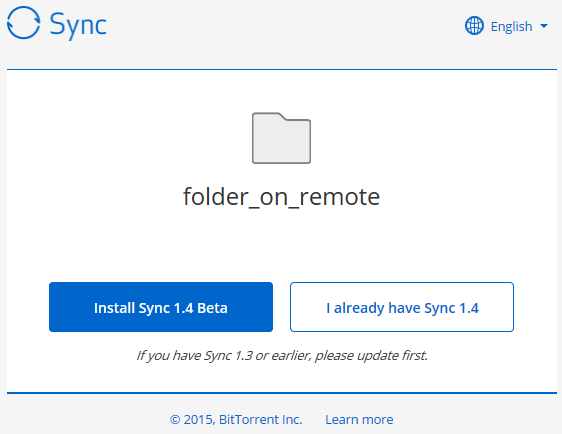}
\caption{A received link can be shortened and still be resolved to a share by the server}
\label{fig:getlink14}
\end{figure}

At a minimum the link must contain the folder, shareID and one time key fields to resolve to a share if entered directly into a browser however removing the version may cause the actual replication to fail if the remote version is incompatible with the version adding the share. An example how this stripped down link resolves is shown in Figure~\ref{fig:getlink14}. Once an option is selected, the share link is converted into a URL that can be opened by the locally installed client if the client satisfies all of the requirements such as version number. 

\begin{figure}
\centering
\includegraphics[width=0.8\textwidth]{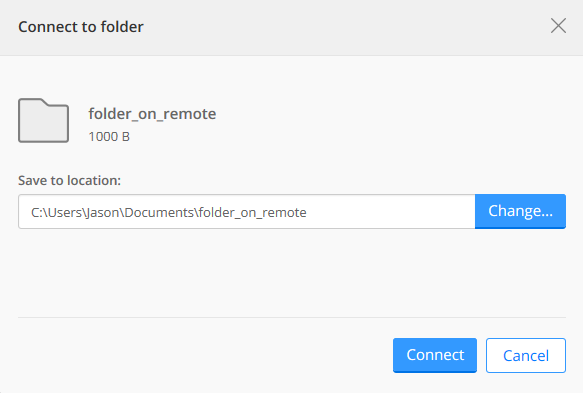}
\caption{A received link can also be added in the section to manually add a share}
\label{fig:clientaddshare14}
\end{figure}

\begin{figure}
\centering
\includegraphics[width=1\textwidth]{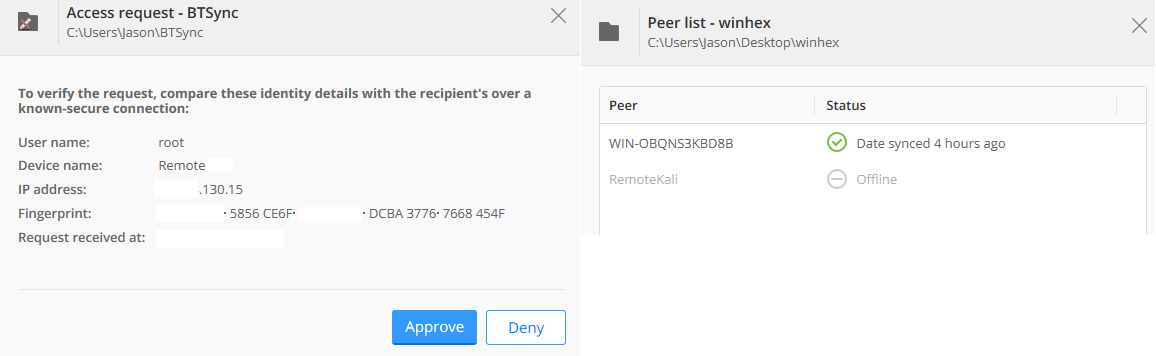}
\caption{Requests for access can be verified (redacted) and share members can be reviewed}
\label{fig:combi1}
\end{figure}

An alternative to opening the link in a web browser is to enter the link in the client itself as if it was a share key, as shown in Figure~\ref{fig:clientaddshare14}. However, if the version is not correct the replication will fail and, if authorisation is required, the request will never be sent to the owner. 

The process of joining a share has also been changed in v1.4 and v2.0. Using the \texttt{x.509} security certificates and public private key pairs stored in the \texttt{sync.dat} file in the \texttt{BitTorrent Sync} folder. Once a host address is retrieved a connection is made and a request for the RO or RW key is sent using the One-Time-Key (i in the optional data) along with the peer's public key generated the first time a link is received or generated. The user and device name set at this time will be the user and device name that the owner will see if they check the identity of the peer requesting access. The device name will also be present in the device list available for each share as can be seen in Figure~\ref{fig:combi1}. Once authorised, the requesting peer receives a copy of the required key encrypted with their public key which they then decrypt and apply to the share on their end of the connection. Once complete the process of synchronisation can begin and the new peer will be registered on the tracker if that option is left enabled.

\subsection{Potential Scenarios Pertinent to Digital Forensic Investigation}
\label{btsync:usecase}

%\subsubsection{Industrial Espionage} Many companies already block BitTorrent traffic with the use of firewall rules configured on their perimeter firewalls. This has the effect of blocking access to tracker servers so torrent files and peers cannot be discovered. With BTSync the user has the option of disabling the requirement for a tracker and to instead use LAN Peer discovery through multicast packets. There is also the option to forgo the use of a tracker altogether and use a configuration setting to allow for ``known peers'', IP:Port pairs that the client contacts directly without having to perform any tracker server lookup. While the user may not be able to transfer files outside the firewall, they may be able to transfer the files to a more vulnerable machine that can bypass the firewall restrictions, such as a system or visitor LAN with unrestricted internet access. 

% prefer the original explanation. It does say all that needs to be said. anything more is a bit long winded in my opinion. However, here it is, maybe you can salvage something useful from it.
\subsubsection{Industrial Espionage} Many companies are aware of the dangers of allowing BitTorrent traffic on their networks. However, quite often corporate IT departments enforce a blocking of the technology through protocol blocking rules on their perimeter firewalls. This has the effect of cutting off any BitTorrent clients installed on the LAN from the outside world. In addition to Deep Packet Inspection (DPI) to investigate the data portion of a network packet passing the inspection point, basic blocking of known torrent tracker sites using firewall rule sets can be used. BTSync does use BitTorrent as the protocol for file transfer but once the transfer session is established using the BTSync protocol all traffic is encrypted using AES and may not be open to inspection by a firewall. It also does not follow the current known patterns that would identify an encrypted BitTorrent stream as the target-source profile is different. Blocking \texttt{t.usyncapp.com} and \texttt{r.usyncapp.com} will stop the tracker and relay options from being used but BTSync can operate quite well without those services. Local peer discovery can use multicast or direct ``known peer" configuration where a known \texttt{IP:Port} combination is used to identify a specific machine allowed to participate in the share. This specificity would negate the issue of multicast packets usually not being routed beyond the current network segment. A scenario where BTSync can be used to transfer files within a LAN would be to transfer data to a machine with lower security protocols in place such as the capability to write to a USB device or perhaps even unmonitored access to the Internet (and the BitTorrent protocol ) through a designated guest LAN.

\subsubsection{Cloudless Backup} By synchronising between two or more machines accessible to the user, data can be stored in multiple locations as a form of backup. The secondary copies of a file would be stored using a read only key so that only changes on the primary system will ever replicated.
A feature of BTSync that is enabled by default but can be disabled in the configuration file, is the use of the \texttt{.SyncArchive} folder that stores a copy of any file deleted or changed for a preset period of time allowing for a form of file recovery or versioning.

\subsubsection{Encrypted Remote P2P Backup} The BitTorrent Sync API \cite{bitsyncapi} adds the functionality to generate an ``encryption secret''. Through the use of encryption secrets, a BTSync user has the ability to remotely store encrypted data, e.g., personal, sensitive or illegal, on one or more remote machines. These remote machines do not have the ability to decrypt the information stored. The data could then be securely wiped off the original machine and easily recovered at a later stage.

\subsubsection{Dead Drop} Due to BTSync's intended use as a file replication utility, it is assumed that a person receiving a copy of a shared directory is aware of the contents of the folder. As a result, no method was included to gather details of the contents of a share before synchronisation. The API \cite{bitsyncapi} introduced this function but only a node configured correctly with an API key will return a folder or file listing when queried. 

\subsubsection{Secure P2P Messaging} For example, the proof of concept found at \texttt{http://missiv.es/}. The application currently operates by saving messages to an ``outbox'' folder that has a read only key shared to the person you want to receive the message. They in turn send you a read only key to their outbox. One to many can be achieved by sharing the read only key with more than one person but no testing has been done with synchronisation timing issues yet and key management may become an issue as a new outbox would be needed for each private conversation required.

\subsubsection{Piracy} -- BitTorrent, like any other P2P technology, was designed for one-to-many distribution of large content and has become almost synonymous with piracy. BTSync was not necessarily intended to be a one-to-many distribution utility. However, it does allow for a group of users to set one another as ``known peers'' so that they can communicate directly through encrypted channels. Websites such as \texttt{http://btsynckeys.com/} have examples of users posting keys publicly and advertising the content as being copyrighted material.

\subsubsection{Serverless Website Hosting} -- This involves the creation of static websites served through a BTSync shared folder. These websites could be directly viewed on each user's local machine. The local copies of the website could receive updates from the webmaster automatically through the synchronisation of the content associated with a read only secret.

\subsubsection{Malicious Software Distribution} -- Due to the lack of any trust level being associated with any publicly shared secret, the synchronised files may contain infected executables.

For each of the above scenarios, an added dimension can be created by the BTSync user: time. Due to the ability to create ``throw away'' or temporary secrets for any piece of content, the timeframe where evidence may be recovered from remote sharing peers might be very short.

%%%%%%%%%%%%%%%%%%%%%%%%%%%%%%%%%%%%%%%%%%%%%%%%%%%%%%%%%%%%%%%%%%%%%%
\section{Related Work}
\label{related}
%%%%%%%%%%%%%%%%%%%%%%%%%%%%%%%%%%%%%%%%%%%%%%%%%%%%%%%%%%%%%%%%%%%%%%

This paper is focused on the network communication protocol employed by BTSync and the investigation thereof. The work presented as part of this paper builds upon the work of Farina et al. \cite{farina2014}, which outlines the forensic analysis of the BTSync client application on a host machine. This paper outlines the procedures for identifying a current or previous install of the BTSync application and the extraction of secrets from gain physical access to a machines hard drive and performing a regular digital forensic investigation on its image. At the time of publication, there are no other academic publications focusing on BTSync. However, seeing as BTSync shares a number of attributes and functionalities with cloud synchronisation services, e.g., Dropbox, Google Drive, etc., and is largely based on the BitTorrent protocol, this section outlines a number of related case studies and investigative techniques for these technologies.

\subsection{BitTorrent Forensics}
\label{btforensics}
%Piracy, Tracker, DHT, Illicit Images of Children
Numerous investigations have been made into identifying the peer information of those involved in BitTorrent swarms. Most of these publications focus on the investigation of the unauthorised distributed of copyrighted material \cite{layton2010investigation}, \cite{scanlon2010week} and \cite{le2010spying}. Depending on the focus of the investigation, peer information may be recorded for a particular piece of material under investigation or a larger landscape view of the peer activity across numerous pieces of content.

\subsection{Client-side Synchronisation Tool Forensics}
\label{clientside}
%Dropbox, Sugarsync, Google Drive

Forensics of cloud storage utilities can prove challenging, as presented by Chung et al. in their 2012 paper \cite{Chung201281}. The difficulty arises because, unless complete local synchronisation has been performed, the data can be stored across various distributed locations. For example, it may only reside in temporary local files, volatile storage (such as the system's RAM) or dispersed across multiple datacentres of the service provider's cloud storage facility. Any digital forensic examination of these systems must pay particular attention to the method of access, usually the Internet browser connecting to the service provider's storage access page (https://www.dropbox.com/login for Dropbox for example). This temporary access serves to highlight the importance of live forensic techniques when investigating a suspect machine as a ``pull out the plug'' anti-forensic technique would not only lose access to any currently opened documents but may also lose any currently stored sessions or other authentication tokens that are stored in RAM. 
%In his paper, Chung describes three main forms of online storage in use by consumers:

%\subsubsection{} Data Storage for large data -- Provided by services such as Amazon S3, Dropbox, Google Drive, etc.
%\subsubsection{} Online only office applications -- This involves an entire productivity suite of tools is accessed in a completely self contained online environment, e.g., Google Docs, Office 365 or Sage Online.
%\subsubsection{} Personal Data -- This involves personal data and preferences being stored in a central repository. For example, Evernote allows users to save notes to a central store and Spotify allows playlists to be stored when accessing their online music catalog.

In 2013, Martini and Choo published the results of a cloud storage forensics investigation on the ownCloud service from both the perspective of the client and the server elements of the service \cite{Martini2013287}. They found that artefacts were found on both the client machine and on the server facilitating the identification of files stored by different users. The module client application was found to store authentication and file metadata relating to files stored on the device itself and on files only stored on the server. Using the client artefacts, the authors were able to decrypt the associated files stored on the server instance.

\subsection{Extension of the Digital Evidence Acquisition Window}
\label{remoteevidence}
In 2014, Scanlon et al., outlined a case study on BTSync whereby the remote recovery of evidence from a BTSync shared folder can enable the recovery of evidence that is no longer accessible on the local machine \cite{scanlon2014leveraging}. This evidence may have been securely deleted, corrupted or overwritten on the local device or viewed (not stored) on a mobile device using the BitTorrent Sync app. The paper outlines a number of entry points from the local machine into the investigation and the remote recovery of such evidence including local and network sources.
%
%Google Drive forensics remnants \cite{Quick2013}
%
%\cite{Quick2013266}
%
%\cite{quick2013digital}

%%%%%%%%%%%%%%%%%%%%%%%%%%%%%%%%%%%%%%%%%%%%%%%%%%%%%%%%%%%%%%%%%%%%%%
\section{BitTorrent Sync Network Protocol Analysis}
\label{protocol}
%%%%%%%%%%%%%%%%%%%%%%%%%%%%%%%%%%%%%%%%%%%%%%%%%%%%%%%%%%%%%%%%%%%%%%

Starting with the beta release of v1.4, BTSync changed its protocol to more closely resemble that of the underlying BitTorrent protocol. In addition to changes to the directory structure and the introduction of public/private key storage for shares, the network traffic profile of the protocol changed dramatically by utilising the Micro Transmission Protocol ($\mu TP$) as outlined in the BitTorrent Extension to Protocol (BEP) 29, which is officially specified here: \texttt{\url{http://www.bittorrent.org/beps/bep_0029.html}}. This protocol was already used by BitTorrent once actual file transfer was initiated but now BTSync has adapted its communications to use $\mu TP$ signalling resulting in a smaller overall usage of bandwidth but a more noticeable footprint. 

Where the initial release of BTSync used custom packets that all started with the header BSYNC[00] or BSync[80], this purely cosmetic identifying header was replaced with the $\mu TP$ DATA version 1 (01) header for all request and transfer packets and STATE (21) was used to perform the same functionality of the original PING used to update peer availability and provide connection details and data.

As with the original $\mu TP$ protocol the connection management packets and headers used by BTSync v1.4 and onwards are:
\begin{description}
\item [SYN :] initiates the two-way $\mu TP$ handshake to establish a connection with the remote peer. This packet has its type indicator set to 4.
\item[STATE :] the most common packet in $\mu TP$, this ``ACK'' replaces the BTSync response to PING and serves as both the keep-alive and the response to the handshake initiation. This packet is identified by the type value of 2.
\item[DATA :] This packet is used to carry messages such as the peer request message sent to the tracker or the peer list sent in response. This packet has a type value of 0.
\item[RST :] as with TCP the RST packet is used to reset the connection in the event of an error in transmission. This has a type identifier set to 3
\item[FIN :] Indicates the end of a connection and is denoted by the type value of 1.
\end{description}

\begin{figure}
\centering
\includegraphics[width=1\textwidth]{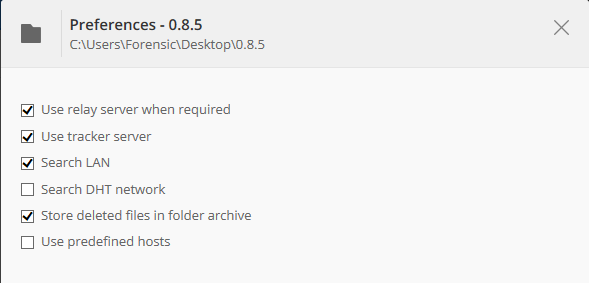}
\caption{A newly created share will have some preferences set by default that can be toggled by the user}
\label{fig:sharepref}
\end{figure}

The $\mu TP$ message headers have a similar layout that is formatted as follows: 
\texttt{Header Type:[0/1/2/3/4]\\Version:[1]\\Extension[00]\\ConnectionID:[AB CD]\\Timestamp:[AB CD EF GH]\\Timestamp Difference[AB CD EF GH]\\Window Size:[AB CD EF GH]\\sequence number:[AB CD]\\Ack Number:[AB CD]}

On provision of a new share several options are enabled automatically by the application as shown in Figure~\ref{fig:sharepref}. These options can be disabled or re-enabled by the user at any time to customise the network behaviour of the local repository being edited. These changes can also be managed through direct editing of the application configuration files. The default behaviour for BTSync is to utilise the tracker server at \texttt{t.usyncapp.com}. The DNS request resolves to three IP addresses: \texttt{54.225.100.8}, \texttt{54.225.92.50} and \texttt{54.225.196.38}. These three IP addresses are servers hosted on Amazon's EC2 cloud service. This is the BTSync tracker server, which facilitates peer discovery for clients looking to synchronise data. One peer request message is sent for each share stored on the local machine and the act of requesting a peer lookup also serves to register the requesting client as a source for that share. 

Packets sent from the client to the tracker server contain registration details and \texttt{get\_peers} message requests (when a new share is created it registers the share with the tracker using a \texttt{get\_peers} packet). A \texttt{get\_peers} packet takes the form of:
\begin{multicols}{2}
\noindent \underline{Version 1.4:}\\
\texttt{Header type: 0}\\
\texttt{d2:la\\6:[6 byte local IP:port]\\2:lp[port integer]\\1:m\\9:get\_peers\\4:peer\\20:[20 byte peer ID]\\5:share\\20:[20 byte ShareID]\\e}
\columnbreak
\\
\underline{Version 2.0:}\\
\texttt{Header type: 0}\\
\texttt{d2:la\\6:[6 byte local IP:Port]\\2:lp [local port integer]\\1:m\\9:get\_peers\\4:peer\\20: [20 byte peer ID]\\5:share\\32:[32 byte ShareID]\\e}
\end{multicols}
\small (where the observed keys are defined in Table \ref{tab:tpacket}).
%\noindent \texttt{BSYNC[00] \\d2:la6:[6-byte IP/Port] \\1:m9:get\_peers \\4:peer20:[20-byte peerID] \\5:share20:[20-byte ShareID] \\e} \\\small (where the observed keys are defined in Table \ref{tab:tpacket}).
\normalsize

\begin{table}
\centering
\caption{Sample Tracker Packet}
\begin{tabular}{|l|p{2.9in}|}
\hline
$\mu TP$ & The $\mu TP$ data header that signifies \\\hline
0x00   & Null \\\hline
d & Start of the dictionary of key:value pairs   \\\hline
2:la  & Local address label identifier which consists of 6-bytes, the first 4 are IP, the last two are port   \\\hline
2:lp  & local port in integer form \\\hline
1:m   & Message label identifier \\\hline
9:get\_peers    & message type value \\\hline
4:peer  & Local peer label \\\hline
20:   & Local PeerID  \\\hline
5:share & Local ShareID label \\\hline
20:    & The 20 character ShareID a transform of the secret used and can be found in the \texttt{.SyncID} file.\\\hline
32: & A share ID based on some transform of the 20 byte ShareID, the local IP address and local port. \\\hline
\end{tabular}

\label{tab:tpacket}
\end{table}

This packet is initially sent to the tracker server via TCP and UDP to test connectivity. If both protocols succeed, UDP is the preferred method of communication. Tracker updates are performed at a rate of once every 600 seconds or if a change is made to the share data, in which case the timer is reset. A separate packet is sent for each share present on the local machine. It is noteworthy that, even when a new share is created, the first packet advertising that share to the server uses a message type of \texttt{get\_peers}. Depending on the bandwidth usage it is entirely possible for a single peer to simultaneously contact and register with multiple tracker server addresses. Each share will have its own Connection ID value in the $\mu TP$ header for that get\_peers packet and each request will prompt a separate type 2 (ACK) response from the tracker server followed by a separate response to the request itself.

The receiving tracker will respond to the requesting client with the same protocol used in the \texttt{get\_peers} message. This has the consequence that if TCP and UDP are successful on the first request, the first response will be a set of duplicate TCP and UDP packet in the form:\\ 
\begin{multicols}{2}
\noindent \underline{Version 1.4 and 2.0:}\\
\texttt{Header type: 0}\\
\texttt{d2:ea\\6:[requester external IP:port]\\1:m\\5:peers\\5:peers\\l[peer list starts]\\e[peer list ends]\\5:share\\20:[20 byte ShareID]\\4:time[timestamp]\\e}
\columnbreak
\\
%\underline{Version 2.0:}\\
%\texttt{Header type: 0}\\
%\texttt{d2:ea\\6:[requester external IP:port]\\1:m\\5:peers\\5:peers\\l[peer list starts]\\e[peer list ends]\\5:share\\20:[20 byte shareID]\\4:time[timestamp]\\e}
\underline{Peer Entry in peer list}\\
\texttt{d\\1:a[address key]\\6:[external IP:Port value]\\2:la[local address key]\\6:[internal IP:Port]\\1:p\\20:[Peer ID]\\e[end of Peer dictionary]}\\
\end{multicols}
 where the observed keys are defined in Table~\ref{tab:tpacket} %and the \texttt{peer list} is in the form:\\
%\underline{Peer List Format}\\
%\texttt{d\\1:a[address key]\\6:[external IP:Port value]\\2:la[local address key]\\6:[internal IP:Port]\\1:p\\20:[Peer ID]\\e[end of Peer dictionary]}\\
The \texttt{peer list} returns an entry for each peer currently in contact with the tracker through \texttt{get\_peer} requests. The current requesting peer will be included in this list so the \texttt{peers} message will always have at least one entry in the peers list.

One unusual feature of the \texttt{peers} response is the inclusion of a peer's local, non-routable, IP address and Port. This is so that, if the local IP matches the local subnet of the requesting peer, the requesting peer can attempt to communicate directly over the LAN using the local address provided. If the tracker server option is disabled then the local client will have to use a different method to find peers local to it. 

\subsection{Local Peer Discovery} 
When the option to search LAN is enabled (the default behaviour) the application will start sending out multicast packets to port 3838 across the LAN. The multicast packets are BTSync bencoded packets with the following format and the keys are further explained in Table \ref{tab:mcastping}.\\
\noindent \texttt{BSYNC[00] \\d1:m4:ping4:peer20:[20-byte Peer ID] \\4:port[i Integer e] \\5:share32:[32-byte content ShareID] e}\\
\normalsize

The format of these packets has not changed since the original pre v1.4 BTSync. Once LAN discovery is enabled the local neighbouring peers will respond to the multicast broadcast with the \texttt{``BSYNC[00]''} TCP packet detailed below.

\begin{table}
\centering
\caption{Multicast Ping Packet}
\begin{tabular}{|l|p{2.9in}|}
\hline
BSYNC  & The BTSync Header  \\\hline
0x00   & Null \\\hline
d & Start of the dictionary of key:value pairs   \\\hline
1:m   & Message label identifier \\\hline
4:PING    &  The message type \\\hline
4:peer  & Local peer label \\\hline
20:   & PeerID of the multicasting Peer  \\\hline
5:share & Local ShareID label \\\hline
32:    & The Share32 ID that matches that used in the v2.0 \texttt{get\_peers} \\\hline
\end{tabular}

\label{tab:mcastping}
\end{table}

Once a peer receives a multicast message that contains a ShareID that it possesses the peer responds with the content: \\\\
\noindent \texttt{BSYNC[00] \\d1:m4:ping4:peer20:[20-byte PeerID] \\4:port[i Integer e] \\5:share20:[20-byte ShareID] e}\\
\normalsize
The keys have the same definitions as those shown in Table \ref{tab:mcastping} with the exception of the ShareID being the more familiar 20 byte version. 

Once the Ping has been sent the peers perform a BTSync session negotiation involving the generation of a nonce value as laid out in Table~\ref{tab:bencoding}. The rest of the synchronisation takes place over TCP IP and the $\mu TP$ traffic runs alongside over UDP. The synchronisation process is signed off with a $\mu TP$ Type 1 (FIN) packet. After this there are regular $\mu TP$ type 2 (STATE) messages to check for changes.

\subsection{BTSync Relay Server} 
\label{BTSyncreq}

When BTSync finds that it needs to communicate directly between two firewalled peers, the application may make use of a relay server. The ``Use Relay Server if required'' option is enabled by default on share creation. The relay server is contacted by a DNS request sent out for \texttt{r.usyncapp.com}, which resolves to the following IP addresses: \texttt{67.215.229.106} and \texttt{67.215.231.242}.

These are the IP addresses of the relay servers contactable on remote port 3000. Each peer contacts the relay server using an outbound connection that should bypass any firewall rule preventing unauthorised inbound connections. Once the server handshake has taken place, the negotiation to set up a secure connection between the two peers begins. The following sequence of events is observable:

\begin{enumerate}
\item{} Peer contacts the relay server to initiate contact with the remote peer.\\
\noindent \texttt{0022 | CounterA | BSYNC 0x000000 [20 byte remote peerID]\\CounterB | peer20 | 20 byte local peerID}
\normalsize

\item{} The relay server responds to the peer using a standard TCP ACK packet\\
%\noindent \texttt{BSYNC[80][remote peerID]\\d1:m4:ping4:peer20:[remote peer ID]\\5:share32:[32-byte ShareID] e}
%\normalsize

%\item{} The peer contacts the relay with a hashmap of the share to indicate which parts are required.\\
%\noindent \texttt{BSYNC[80][remote peerID][non-bencoded data including a hashmap] e}
%\normalsize
%
%\item{} The server responds to the peer with a hashmap of the remote share to conclude the exchange of data availability.\\
%\noindent \texttt{BSYNC[80][remote peerID][non-bencoded data including hashmap] e}
%\normalsize

\item{} The peer contacts the server to arrange transfer of the data and to supply the nonce for encrypted traffic and provide a status ID.\\
\noindent \texttt{0066 | CounterA | BSYNC 0x00(4) :d5:nonce16:[nonce value for key share]5:share20:[20 byte shareID]e}
\normalsize\\

\item{} The relay contacts the peer to initiate the session counters\\
\noindent \texttt{0022 | CounterA | [20 byte remote peerID]\\remote Peer IP:Port | Counter B }
\normalsize

\item{} The relay server Confirms the SID status and supplies the remote nonce to complete the bridge for encrypted data transfer\\
\noindent \texttt{0022 | CounterB | remote peerID | 0066 | remote Peer ID | CounterA \\
 BSYNC 00x4 | :d5:Nonce16:[nonce value]5:share20:[20 byte ShareID]e}
\normalsize

\item{} The Relay server contacts the local peer to deliver the remote public key\\

\item {} the local peer delivers its public key to the relay server\\

\item{} Encrypted bidirectional traffic transfer commences with the relay server acting as the router delivering packets to each peer.
\end{enumerate}

\subsection{BTSync Data Transfer}
\label{InternetSync}

The transfer of data during a BTSync synchronisation operates in a similar fashion as a regular BitTorrent download as described in Section \ref{downloading} above. A unique magnet URI is created for each file contained within the shared folder and this is used for requesting chunks of the entire file from known peers sharing this content.

\subsection{Differentiation from Regular BitTorrent Traffic}
\label{differentiation}

While much of the network topology of BTSync is shared with regular BitTorrent, the request and response packets differ from those employed by regular BitTorrent file-sharing traffic. The most obvious addition is the \texttt{BSYNC} header attached to each datagram transmitted on the network. In addition, the introduction of $\mu TP$ causes increased volume of traffic recognisable even though $\mu TP$ results in lower overall bandwidth usage. Besides that addition, the active peer list that is returned also contains additional information over the regular BitTorrent file-sharing protocol: namely the inclusion of the local IP:port address pairs for each peer. From an investigative perspective, this extra information could prove useful in identifying the particular machine involved in the BTSync network as opposed to merely resolving the WAN IP address back to a router with potentially hundreds of LAN users. The local DHCP records could be used to resolve the MAC address (and often the hostname) of the individual machine identified during the network investigation.

In addition to the regular BitTorrent peer discovery methods outline in Section \ref{discovery} above, BTSync also allows the user to manually add known IP addresses to the local cache of peers. BTSync facilitates this through the option to add ``Predefined Hosts'' to the configuration or application options. These are hardcoded IP address and port entries that are saved in order of preference. BTSync will contact these peers directly, without any requirement for a multicast (LPD) or sending a \texttt{get\_peers} request to an online tracker.

%%%%%%%%%%%%%%%%%%%%%%%%%%%%%%%%%%%%%%%%%%%%%%%%%%%%%%%%%%%%%%%
\section{Investigation Methodology}
\label{methodology}
%%%%%%%%%%%%%%%%%%%%%%%%%%%%%%%%%%%%%%%%%%%%%%%%%%%%%%%%%%%%%%%

\begin{figure}
\centering
\includegraphics[width=0.5\textwidth]{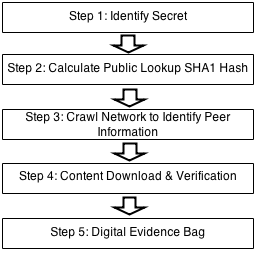}
\caption{Steps Involved in Performing a BTSync Network Investigation}
\label{fig:methodology}
\end{figure}

This section outlines a reproducible methodology for the network investigation methodology. Depending on which of the scenarios outlined above, the methodology may branch according to what the desired outcome will be. Figure~\ref{fig:methodology} outlines the five steps involved in the investigative process (each of these steps are described in greater detail below).

\subsection{Identification of Content}
\label{identification}

Depending on the scenario that motivates the BTSync network investigation, there are a number of avenues that the forensic investigator may find secrets (and corresponding hash values) needed for investigation:
\subsubsection{Web Discovery} -- As soon as BTSync was released as a public alpha, publicly accessible sharing secrets started to appear online. Two ``subreddits'' appeared on Reddit \cite{reddit} and numerous websites and blogs were created to set up an online ``dead drop'' secret share, for example \texttt{http://www.12char.com} and \texttt{http://www.btsynckeys.com}. It is also feasible that an investigator could come across an online community that shares secrets in a private forum for the purposes of trading data and material without 3rd party involvement. Keys to shares discovered in this manner that possess a timestamp component will need to be checked to determine if the link has expired or not.

\subsubsection{Local Discovery} -- An investigator could, in the course of an investigation find evidence of BTSync having been used to transfer material to the suspect machine. This could be that BTSync installed and the folder listed in the list of shares stored in the configuration file, webUI or the BTSync hidden \texttt{.Sync} folder. BTSync log files (/.sync/sync.log), or, if BTSync is not present (uninstalled) there could still be \texttt{.SyncID} files remaining in folders that were synchronised from remote peers. A hexdump of the \texttt{.SyncID} file or, more conveniently, the names of the \texttt{*.db} files found in the \texttt{.Sync} folder will give the SHA1 encoded share ID that the investigator needs to find other peers actively sharing that content

\subsubsection{LAN traffic} -- Many companies configure their edge firewalls to block torrent traffic for the general users. If the company uses torrent for some other business purpose it will usually be accounted for and allowed from or to a particular server or subnet. However, BTSync allows for all external communicate beyond the LAN to be turned off (in the configuration file or in the settings dialogue the options for ``Use DHT'', ``Use Tracker'' and ``Use Relay Server'' can be disabled) leaving only the settings for LAN discovery or known peers. A security review of the router logs may find active torrent traffic within the LAN or system admins may discover evidence of torrent applications run.

\subsection{Identification of Lookup Hash}
\label{hash}
Requesting a list of peers through any of the peer discovery methods outlined above requires a unique lookup hash. This hash is used by the tracker, DHT, PEX and LPD in the association of know peers to a particular piece of content.

\subsection{Crawl the Network to Identify Peer Information}
\label{crawl}
Each of the peer discovery methods outlined above should be queried for a list of known active nodes sharing that content. Due to the user configurable nature over which services are enabled in the BTSync client, to ensure complete node enumeration/identification, the results from each of the peer discovery methods should be combined to form the final result of collected information.

\subsection{Downloading and Verification of Content}
\label{verification}
Depending on the scenario being investigated, it may be necessary to download a copy of the content stored remotely for investigation or verification. In order to accomplish this, a regular BitTorrent download can be started for each of the files contained within the shared folder. If the investigation's goal is to attempt to recreate content deliberately deleted off a suspect's machine, the data can only be entirely recovered if there is a complete copy of the data stored remotely. However, this does not mean that any single node needs to have 100\% of the content. The original data can be recombined so long as a complete copy exists split among the distributed nodes actively sharing the content. An obstacle to this stage of the investigation would be the use of limited use keys. The link descriptor for a key has no component to indicate a restricted number of uses. A further obstacle would be the option to require authorisation before a peer can access a share. This is unlikely to be the case for links discovered on a public platform.

%\subsection{Digital Evidence Bag}
%\label{DEB}
%Once the required information is gathered, the resulting data and all associated metadata (peer information, file sizes, hash values, etc.) should be gathered together into a suitable digital evidence bag. For verifiable reproduction of the results achieved, a copy of the network stream created during the investigation should be stored as part of this digital evidence bag, as outlined in detail by Scanlon et al. \cite{scanlon2014digital}.

%
%Depending on the scenario that motivates the BTSync network investigation, there are a number of avenues that the forensic investigator may find secrets (and corresponding hash values) needed for investigation
%\begin{enumerate}
%\item Blah -- 
%\item Blah -- 
%\end{enumerate}

%%%%%%%%%%%%%%%%%%%%%%%%%%%%%%%%%%%%%%%%%%%%%%%%%%%%%%%%%%%%%%%
\section{Proof of Concept}
\label{proofofconcept}
%%%%%%%%%%%%%%%%%%%%%%%%%%%%%%%%%%%%%%%%%%%%%%%%%%%%%%%%%%%%%%%

In order to begin proof of concept testing for the investigation methodology, a bespoke BTSync crawling application was first designed and developed. This application was built to emulate regular BTSync client usage, as outlined above, and recorded the necessary results for analysis. 

\subsection{Overview}
\label{investigationoverview}
To demonstrate the functionality of the application, an investigation was conducted on a known publicly accessible BTSync secret. One of the public BTSync online secret sharing sites was used (\texttt{http://www.btsynckeys.com/}) to acquire a secret likely to have active peers sharing the corresponding content. The secret selected was advertised with the description ``\texttt{45 GB Movie Collection [Movies] [R]}'' and the read-only secret \texttt{BKV273YUFMWILMESLRDVLI5NHMWO3OCS7} was supplied. It is important to note that there is no certainty that the description accurately advertises the content within the share. There is no method of verifying any of the containing shared content until the syncing process begins and temporary files are created in the shared folder. Even at that point, the user can merely see the file names of the content once the download/synchronisation process has begun.

\subsection{Results}
\label{results}

As part of the peer identification process a number of active peers were returned to the investigative application. These peers were recorded for later analysis. During the first snapshot taken for this investigation, 21 peers were identified as sharing the specific content and 20 were identified on the second. A snapshot accounts for all of the peers identified sharing the specific content at the same instance in time.

\begin{figure}[t]
\centering
\includegraphics[width=0.8\textwidth]{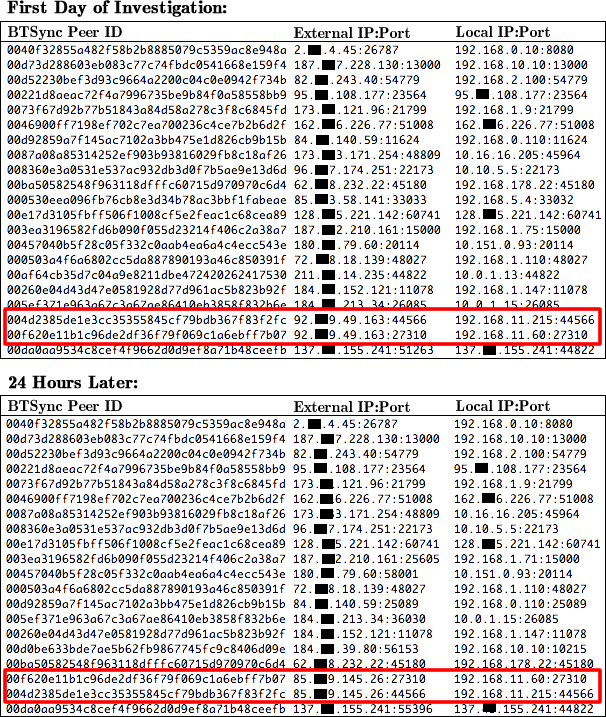}
\caption{Daily Snapshot Comparison for Investigated Secret (Public IP Addresses Partially Redacted)}
\label{comparison}
\end{figure}

Two peers (differentiated by PeerID) of particular interest are listed as the second and third last peers in both tables in Figure~\ref{comparison} (highlighted in red). Comparing their peer ID and local IP:Port address pairing, it is clear that these two peers are referring to the same individual node. Between the two snapshots taken of this shared content, their IP address changed from one IP address range to another. However, both of these IP address ranges are associated with the ISP ``Telefonica'' in the same postal zip code in Berlin, Germany (data gathered from Maxmind \cite{maxmind}). This information indicates an ISP level IP address reallocation sometime between the two snapshots as opposed to the use of a VPN or other IP address masking system. The two peers share the same external IP address but have different external ports and local IP:port pairs indicating that the BTSync install on these nodes are accessing the Internet through a router employing Network Address Translation (NAT).

\subsection{Churn Rate}
\label{churn}
While the example investigation outlined as part of this paper focuses on a single secret over a 24 hour window, the low churn rate of just 7\% remains interesting. Most P2P networks experience a high turnover of peers \cite{herrera2007modeling}; following the assumption that most users are active on the network while downloading some content and disconnect upon completion. BTSync is designed to be a tool that functions in a similar manner to cloud file synchronisation services like Dropbox or Google Drive. These tools largely operate on an ``install and forget'' approach whereby synchronisation and updating between the cloud and potentially multiple client machines does not require any direct user input. BTSync uses a similar approach and as a result, low churn rates would be expected.

\subsection{Geolocation}
\label{geolocation}
\begin{figure}
\centering
\includegraphics[width=0.8\textwidth]{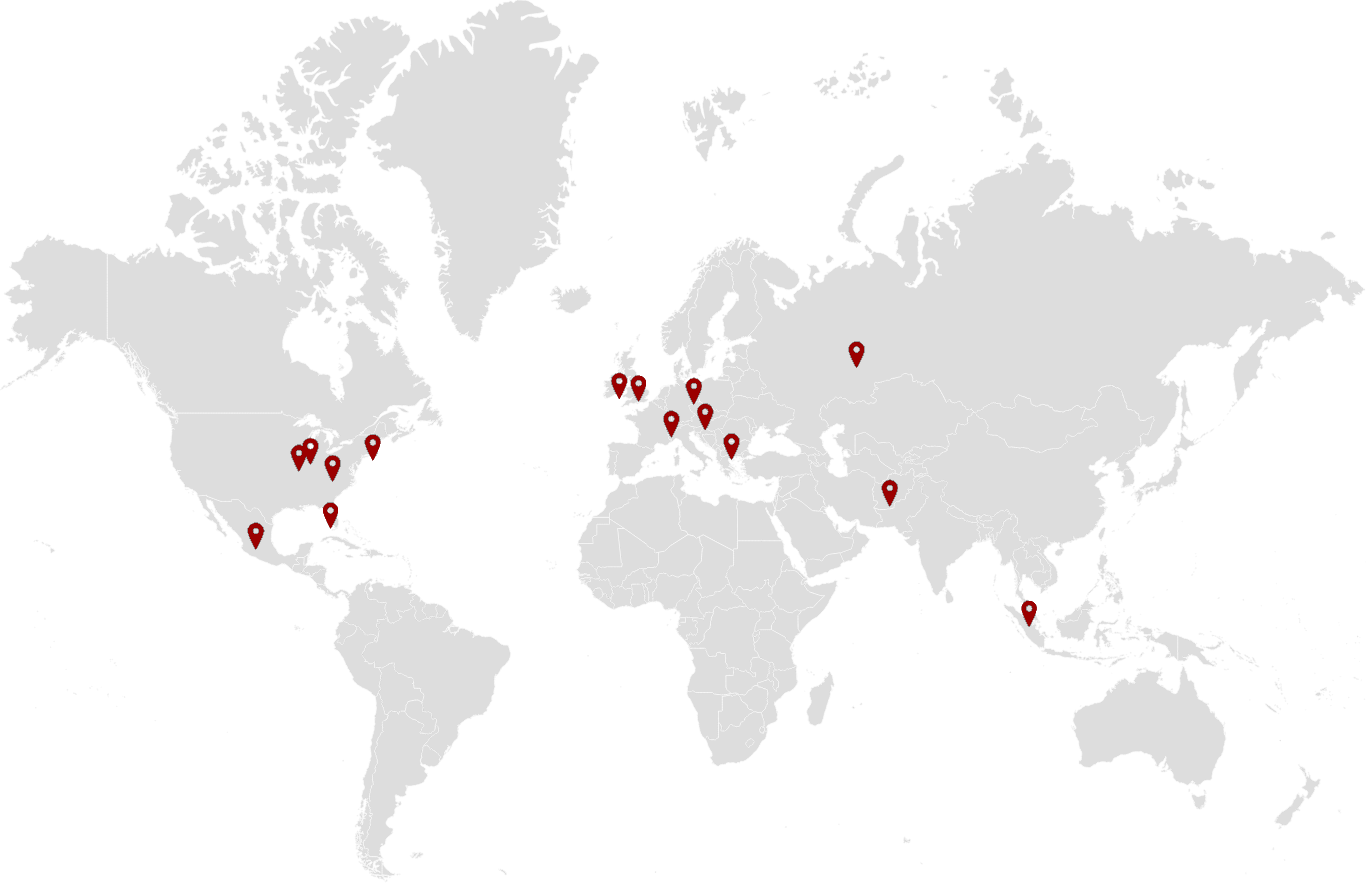}
\caption{Geolocation of Discovered IP Addresses}
\label{fig:geolocation}
\end{figure}

Figure~\ref{fig:geolocation} shows the geographic distribution of the peers identified as part of the investigation. While the total number of peers identified with this proof of concept investigation is quite low, the data remains consistent with regular BitTorrent investigative results \cite{scanlon2010week} with North America and Europe being the most popular continents involved.

\section{Example Investigation}
\label{investigation}
In late August 2014, the iCloud accounts of numerous celebrities were hacked and compromising photos and videos were posted online without their consent in what has gained notoriety in the media and among Internet users as "The Fappening" \cite{ibtimes} or "Celebgate" \cite{muth2015googlestroika}. The comprised photos spread across the globe with the help of Internet forums, such as \texttt{htpp://4chan.org} and \texttt{http://reddit.com}. At the time, there was concerns that iCloud itself had been hacked and these leaks were merely a subset of the information stolen of Apple's servers, however an investigation into the attack found that the passwords were cracked for specific accounts \cite{ibtimes}.

\subsection{Entry Point}
The entry point to this investigation first involved verifying that this content was being shared using BTSync. On the public BTSync secret sharing ``subreddit'' \texttt{http://reddit.com/r/btsecrets}, a number of public read-only secrets were shared containing collections of the leaked content. For the purposes of this investigation, one shared leaked content was investigated using the aforementioned BTSync investigative application. The secret investigated was \texttt{bb63eb5b61969956e71273026f00a1deca464413}. The investigation took place one week after the leak occurred.

\begin{figure}[!t]
\centering
\includegraphics[width=\textwidth]{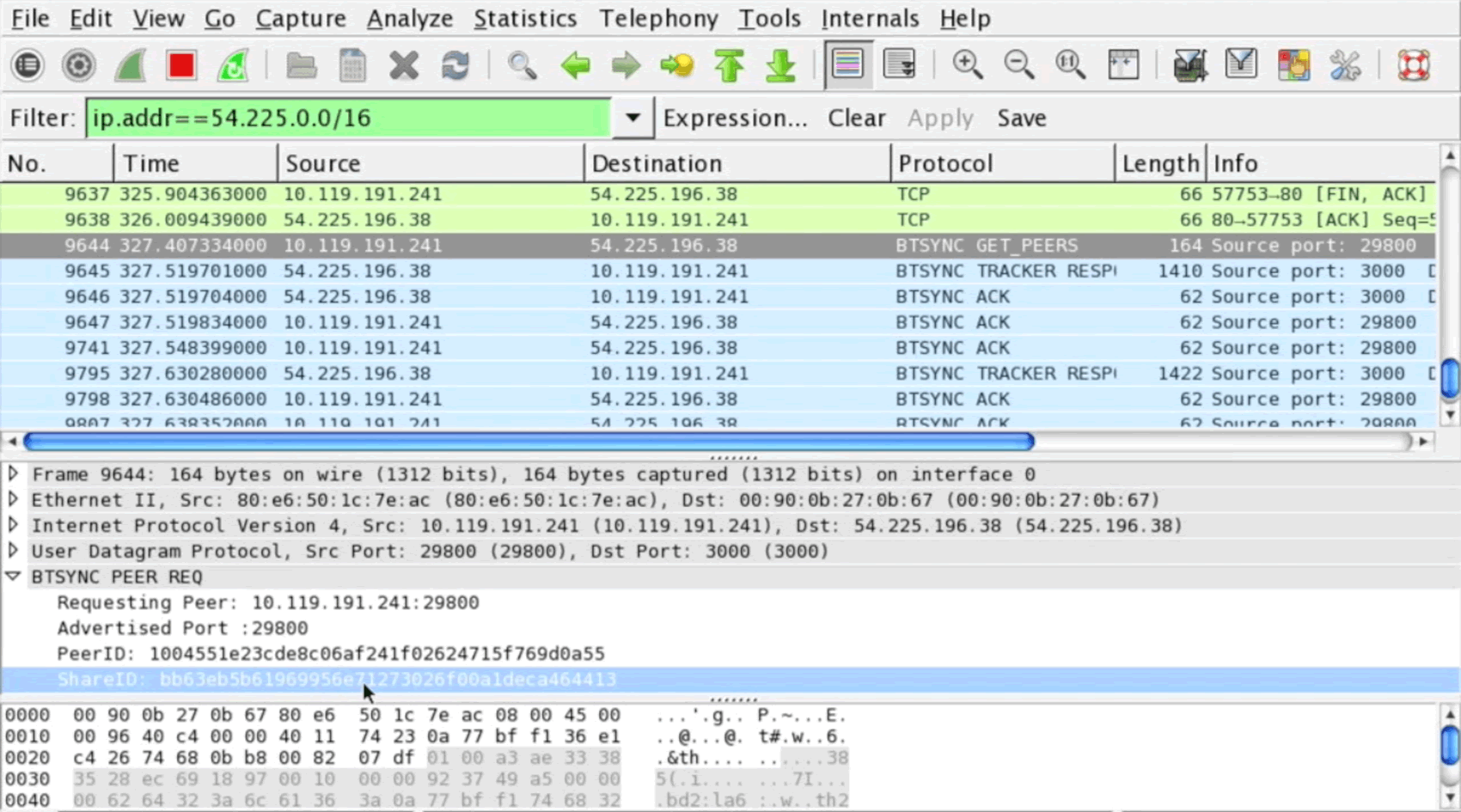}
\caption{Network-based Entry Point into Investigation (\texttt{ShareID} Highlighted in Blue}
\label{fig:exentry}
\end{figure}

A BTSync dissector for Wireshark was developed\footnote{Wireshark Dissector is downloadable from \url{http://www.markscanlon.co/bittorrent-sync}} to expedite the network analysis process. This dissector can identify the various packets pertinent to the decentralised service in the Wireshark traffic capture, as can be seen in Figure~\ref{fig:exentry}. Using the gathered \texttt{ShareID} from the network traffic, the investigative application was launched and the \texttt{ShareID} supplied. 

\subsection{Peer Discovery}
\begin{figure}[!h]
\centering
\includegraphics[width=\textwidth]{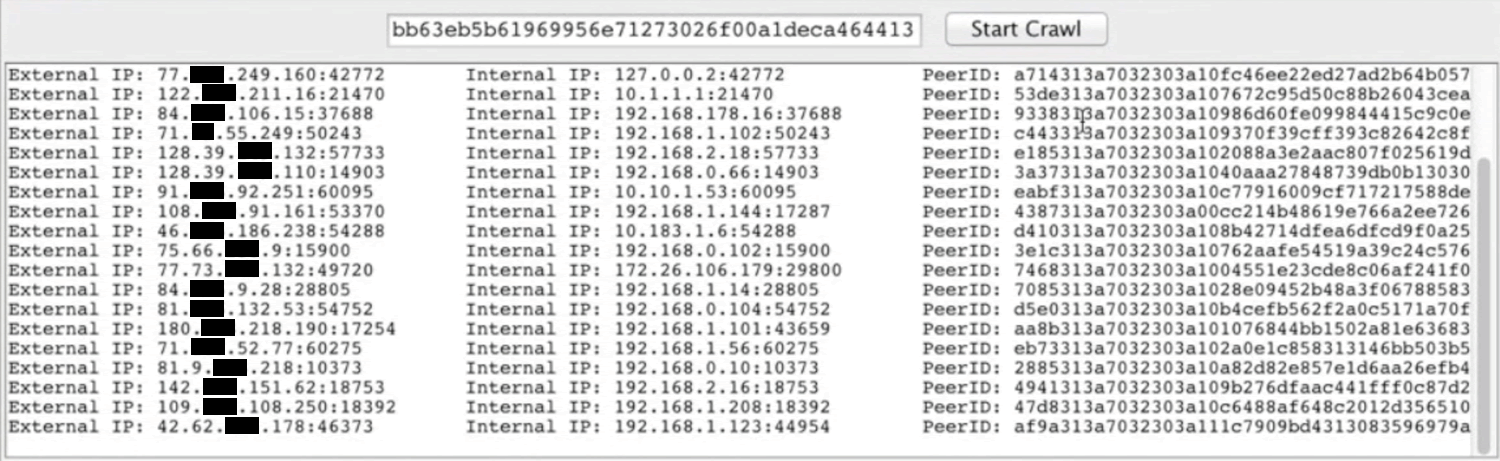}
\caption{IP Addresses Discovered Sharing the Content}
\label{fig:exips}
\end{figure}

Using the gathered \texttt{ShareID}, the application was able to gather information about each of the peers sharing the content, as can be seen in Figure~\ref{fig:exips} using each of the peer discovery methods outlined above. 

\subsection{Geolocation}
\begin{figure}[!h]
\centering
\includegraphics[width=0.8\textwidth]{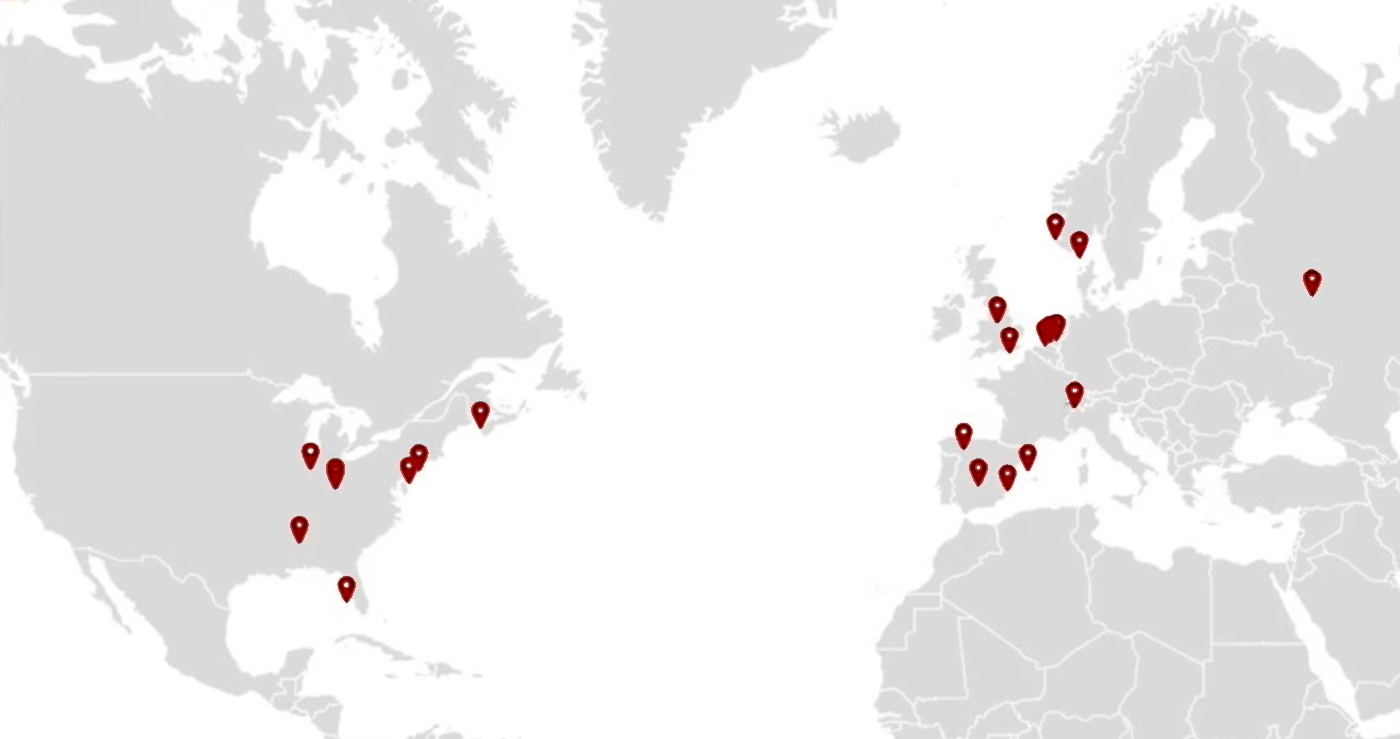}
\caption{Geolocation of Discovered Peers}
\label{fig:exgeo}
\end{figure}

The IP addresses detected during the investigation were geolocated and found to be located in North America and Europe, as can be seen in Figure~\ref{fig:exgeo}.

\subsection{Data Recoverable from Remote Peers}
\begin{figure}[!h]
\centering
\includegraphics[width=0.8\textwidth]{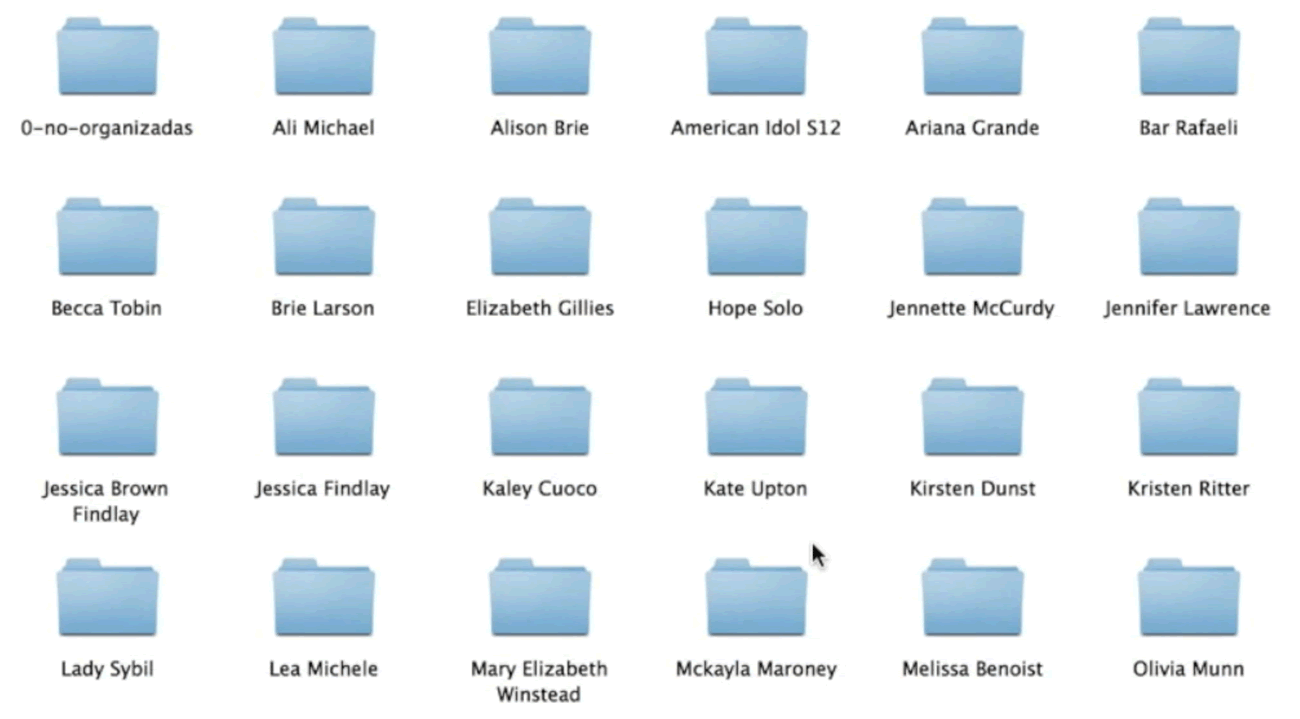}
\caption{Evidence Recovery from Remote Peers}
\label{fig:exevidence}
\end{figure}

Some of the evidence recoverable from remote peers in this particular BTSync share can be seen in Figure~\ref{fig:exevidence}. The version of the BTSync available at the time of the investigation (v1.4), did not have selective sync functionality. As a result, each member of the secret must download all of the shared content. This limitation of a lack of selective syncing means that each peer identified will eventually have all of the content in the share. This feature makes evidence recovery from such popular shares more performant for digital investigators as each node is a potential source of the pertinent evidence. With the advent of v2.0 of the application, selective sync means that each peer must be communicated with individually to identify which active machines identified has what data.

\section{Conclusion}
\label{conclusion}
This paper documented the protocol used in BitTorrent Sync during the discovery of peers and the synchronisation of data. While BTSync is not necessarily intended to replace BitTorrent as a file dissemination utility, it will likely be used for this purpose. This is already facilitated though websites providing shared secrets, e.g., Reddit \cite{reddit}, etc., as a form of dead-drop. The developers describe the tool as an end-to-end encrypted method of transferring files without the use of a third party staging area, which ensures that the content and personal details remain hidden from unauthorised access. Analysis of the network communication procedure produced unique identifiable information on peers including their unique PeerID, their external and local IP addresses and port numbers. In combination with traditional digital forensic methods, once a secret is identified, it is possible to discover other nodes on the network who are also sharing this data. Deleted data from a local shared folder could be downloaded from the network and recombined for forensic investigation. From an investigative perspective, the decentralised nature of BTSync will always leave an avenue of gathering information and identifying nodes sharing particular content open to the forensic investigator.
\section*{References}

\bibliography{mybibfile}

\end{document}